\documentclass[epj]{svjour}
\usepackage{amssymb}
\usepackage{latexsym}
\usepackage[dvipdfmx]{graphicx}
\begin{document}
\title{
Criticality of the low-frequency conductivity for the 
bilayer quantum Heisenberg model
}
\author{Yoshihiro Nishiyama  
}                     
\offprints{}          
\institute{Department of Physics, Faculty of Science,
Okayama University, Okayama 700-8530, Japan}
\date{Received: date / Revised version: date}
%
\abstract{
The
criticality of the low-frequency conductivity 
for the
bilayer quantum Heisenberg model was investigated numerically.
The dynamical conductivity
(associated with the 
O$(3)$ symmetry)
displays the
inductor $\sigma (\omega) =(i\omega L)^{-1}$ and 
capacitor $i \omega C$
behaviors
for the ordered
and disordered phases, respectively.
Both constants, $C$ and $L$,
have the same scaling dimension as 
that of the reciprocal
paramagnetic gap $\Delta^{-1}$.
Then,
there arose a question
to fix the set of
critical amplitude ratios among them.
So far, the O$(2)$ case 
has been investigated 
in the context of
the boson-vortex duality.
In this paper,
we employ
the 
exact diagonalization method,
which enables us to calculate
the paramagnetic gap $\Delta$ directly.
Thereby,
the 
set of critical amplitude ratios
as to $C$, $L$ and $\Delta$
are estimated with the finite-size-scaling  
analysis for
the cluster with $N \le 34$ spins.
} 
\maketitle

\section{\label{section1}Introduction}

\begin{figure}
\resizebox{0.5\textwidth}{!}{%
\includegraphics{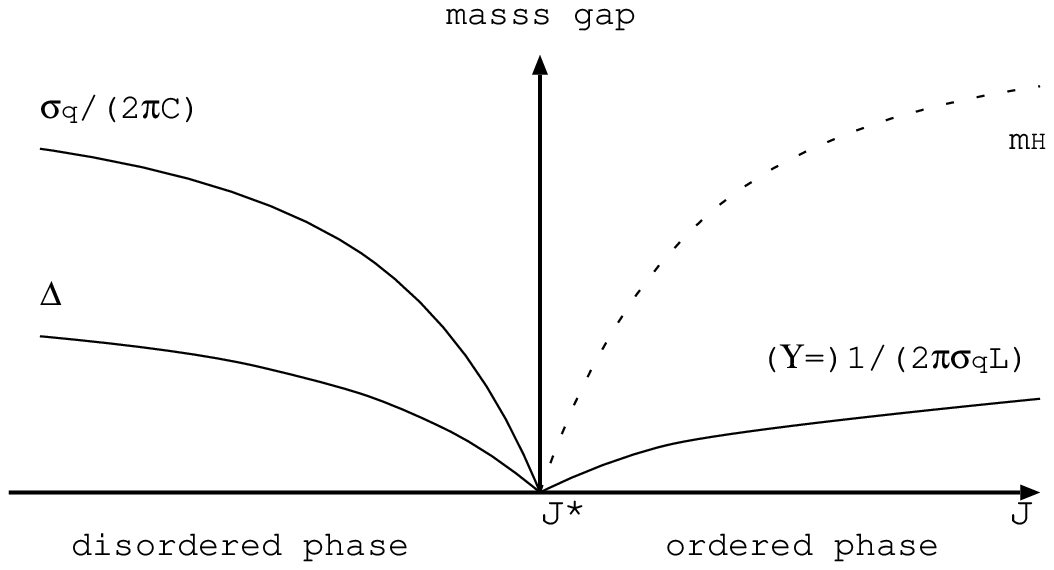}  }%
\caption{
\label{figure1}
A schematic drawing of the transport and spectral properties
is presented
for the $(2+1)$-dimensional Heisenberg model
in both ordered 
($J>J^*$)
and 
disordered
($J<J^*$)
phases.
Here, the symbols,
$L$, $C$, $\Delta$, $m_H$, $\Upsilon$ and $\sigma_q$
denote
inductance, capacitance, paramagnetic gap,
Higgs mass, helicity modulus, and quantum conductance,
respectively.
The scaling dimensions of
$L^{-1}$,
$C^{-1}$, $\Delta$, and $m_H$,
are identical,
and the amplitude ratios among them
make sense.
Particularly,
the amplitude ratio $m_H/\Delta$ has been
scrutinized rather extensively 
\cite{Gazit13,Lohofer15,Rose15,Katan15,Nishiyama16};
the Higgs excitation may have a short life time
\cite{Rancon14}.
}
\end{figure}

For the O$(N)$-symmetric $(2+1)$-dimensional
system,
the low-frequency conductivity
(associated with the O$(N)$ symmetry group)
exhibits the
inductor $\sigma(\omega)=(i \omega L)^{-1}$
and capacitor 
$i\omega C$ behaviors
in the 
ordered
and disordered 
 phases,
respectively \cite{Gazit13}.
In Fig. \ref{figure1},
a schematic drawing of $L^{-1}$ and $C^{-1}$
is presented for both ordered 
($J>J^*$)
and 
disordered
($J<J^*$)
phases;
here, the symbols, $\Upsilon$ and $\sigma_q$,
denote
the helicity modulus
and the quantum conductance ($\sigma_q=q^2/h$),
respectively.
A key ingredient is that
the conductivity in two (spatial) dimensions is scale-invariant,
and both constants,
$L^{-1}$ and $C^{-1}$,
have the same scaling dimension as that of
the paramagnetic gap $\Delta$;
note that the angular velocity $\omega$
has the same scaling dimension as that of the energy gap 
(reciprocal correlation length).
Then,
there arose a question to fix
the set of
amplitude ratios among $L^{-1}$, $C^{-1}$ and $\Delta$.
These parameters
govern the low-energy physics
for both transport and spectral properties \cite{Chubukov94}.
For generic values of $N=2,3,\dots$,
these amplitude ratios were estimated
with the non-perturbative renormalization-group method
\cite{Rose17}; an overview 
is presented afterward.
In Fig. \ref{figure1},
the Higgs mass gap $m_H$ 
is shown
as well.
The critical amplitude ratio $m_H/\Delta$ 
has been investigated
rather extensively 
\cite{Gazit13,Lohofer15,Rose15,Katan15,Nishiyama16}.
The Higgs particle may have a short life time for $N \ge 3$
\cite{Rancon14};
the extended symmetry group O$(N)$
leads to enhanced
Goldstone-mode-mediated decay of the Higgs particle.

The underlying physics behind
the amplitude ratio
$C/L$ would be elucidated by the 
duality theory for O$(2)$ \cite{Stone78,Fisher89,Wen90}.
The case O$(2)$ is 
relevant to the superfluid-insulator transition.
According to the duality theory,
the boson conductivity
$\sigma$ and its dual one (vortex conductivity)
$\sigma_v$ satisfy the reciprocal relation
$\sigma \sigma_v=q^2/h^2$,
resulting in
 the contrasting behaviors
between the superfluid and insulator phases
for the transport properties;
see Fig. \ref{figure1}.
Correspondingly,
the
superfluid and Mott-insulator phases are characterized
by the superfluid density 
$\rho_{s}(=\Upsilon)$
and the vortex-condensation stiffness $\rho_v$,
respectively.  
These constants 
$\rho_{s,v}$
are related to the reactance as
$\rho_s=\hbar/(2\pi\sigma_q L)$
and
$\rho_v=\sigma_q\hbar/(2 \pi C)$,
respectively \cite{Gazit13}.
Therefore,
the amplitude ratio
$\rho_s/\rho_v=C/(L\sigma_q^2)$
admits
a ``quantitative measure''
\cite{Gazit14} of deviation from self-duality.
As a matter of fact,
the renormalization group method \cite{Rose17} yields 
$\rho_s/\rho_v=0.210$ ($N=2$),    
which indicates marked deviation from self-duality 
($\rho_s/\rho_v = 1$).
Although 
the duality idea does not apply to the O$(3)$ case,
the amplitude ratio $C/L$ still makes sense,
and worth considering \cite{Rose17}.
Experimentally \cite{Corson99,Crane07,Sherson10},
the vortex-condensate stiffness $\rho_v$
(equivalently, $C$)
is an observable quantity
\cite{Gazit14},
and hence,
the amplitude ratio 
$\rho_s/\rho_v$
is not a mere theoretical concept.

In this paper, we devote ourselves to the case 
O$(3)$.
For that purpose,
we consider the bilayer Heisenberg model (\ref{Hamiltonian}),
which exhibits \cite{Troyer98}
the phase  transition belonging to the O$(3)$-universality class
\cite{Hasenbusch01,Campostrini02}.
We employed
the exact diagonalization method,
which
allows us to calculate the dynamical quantities such as the paramagnetic gap $\Delta$
without resorting to the inverse Laplace transformation
(numerical analytical continuation)
\cite{Gazit13}.
The O$(3)$ case has been investigated with the 
non-perturbative renormalization group 
\cite{Rose17,Rose15,Rancon13} and
Monte Carlo \cite{Gazit13}
methods.

The Hamiltonian for the bilayer Heisenberg model
is given by
\begin{equation}
\label{Hamiltonian}
{\cal H}=
-J \sum_{a=1}^2 \sum_{\langle ij \rangle} {\bf S}_{ai} \cdot {\bf S}_{aj}
-J_{2} \sum_{a=1}^2 \sum_{\langle \langle ij \rangle \rangle} 
     {\bf S}_{ai} \cdot {\bf S}_{aj}
+ J'\sum_i^{N/2} {\bf S}_{1i} \cdot {\bf S}_{2i}
  .
\end{equation}
Here, the quantum spin 
${\bf S}_{ai}$
is placed at each square-lattice
point $i=1,2,\dots,N/2$
within
the layer specified by $a=1,2$.
The summations,
$\sum_{\langle ij \rangle}$
and 
$\sum_{\langle \langle ij \rangle \rangle}$,
run over all possible 
nearest-neighbor and
next-nearest-neighbor
pairs,
$\langle ij\rangle$ and
$\langle \langle ij \rangle \rangle$, respectively,
within each layer.
The parameters $J$ and $J_2$
are the corresponding coupling constants.
The variable $J'$
denotes
the inter-layer antiferromagnetic interaction,
which stabilizes the paramagnetic phase.
According to the Monte Carlo simulation \cite{Troyer98},
a critical point 
\begin{equation}
\label{critical_point}
(J^*,J_2^*,J'^*)=(0.435,0,1)
,
\end{equation}
was found.
Our simulation was performed around this critical point.

It has to be mentioned that the conductivity for the 
Heisenberg model
has been investigated extensively
in the context of the spintronics
\cite{Sentef07,Pardini08,Kubo13,Chen13}.
In this paper, we dwell on
the
criticality of the conductivity
for both ordered and disordered phases.
For that purpose,
we extended the Heisenberg model to the bilayer one (\ref{Hamiltonian})
so as to realize the phase transition
by tuning the redundant coupling constants $(J,J_2)$.


The rest of this paper is organized as follows.
In 
Sec. \ref{section2}, we present the simulation results.
Technical details are presented in Appendix.
In Sec.
\ref{section3},
we address the summary and discussions.

\section{\label{section2}
Numerical results}

In this section, we analyze the 
amplitude ratios
as to
$L$, $C$ and $\Delta$.
For that purpose,
we simulate
the bilayer Heisenberg model 
(\ref{Hamiltonian})
by means of the exact diagonalization method
under 
the settings
$\hbar=q^2=1$.
We implemented the screw-boundary condition
\cite{Novotny90}
so as to
treat a variety of system sizes
$N=18,20,\dots$ systematically.
The algorithm is based on 
the formula (A.1) of Ref.
\cite{Nishiyama16};
however, in order to cope with
the 
next-nearest-neighbor interaction $J_2$,
a number of extensions are required 
as explicated in Appendix.
The linear dimension $\ell$ of the cluster
is given by $\ell =\sqrt{N/2}$,
because the $N/2$ spins constitute a rectangular layer,

\subsection{\label{section2_1}
Amplitude ratio $\Upsilon / \Delta(=\hbar/2\pi\sigma_q L \Delta)$}

In this section, we estimate the amplitude ratio 
$\Upsilon /\Delta(=\hbar/2\pi\sigma_q L \Delta)$.
We surveyed the
interaction subspace
\begin{equation}
(J,J_2,J')=(J^*,J_2^*,J'^*)+(\delta J, \delta J_2, 0)
,
\end{equation}
with
the critical point
$(J^*,J_2^*,J'^*)=(0.435,0,1)$ \cite{Troyer98}
and $\delta J_2=2 \delta J$.
Within this interaction subspace,
the ratio $\Upsilon /\Delta$ 
turned out to
exhibit a stable plateau
for a considerably wide range of
$\delta J$;
see Fig. \ref{figure3} mentioned afterward.

\begin{figure}
\resizebox{0.5\textwidth}{!}{%
\includegraphics{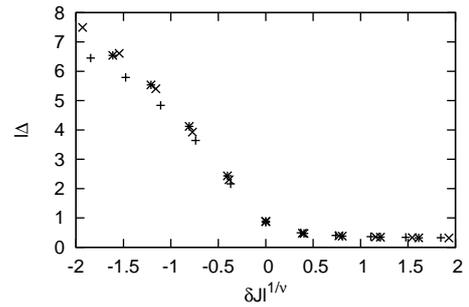}  }%
\caption{
\label{figure2}
The scaling plot of the paramagnetic gap,
$\delta J \ell^{1/\nu}$-$\ell \Delta$,
is presented
for the system sizes,
$N=30$ ($+$),
$32$ ($\times$), and
$34$ ($*$).
The scaling parameter 
(correlation-length critical exponent)
$\nu=0.7112$ 
is taken from the existing literatures 
\cite{Hasenbusch01,Campostrini02}
(3D Heisenberg universality class);
hence, there is no adjustable fitting parameter 
involved in the scaling analysis.
The paramagnetic gap opens in the disordered phase
 ($\delta J<0$).
}
\end{figure}

\begin{figure}
\resizebox{0.5\textwidth}{!}{%
\includegraphics{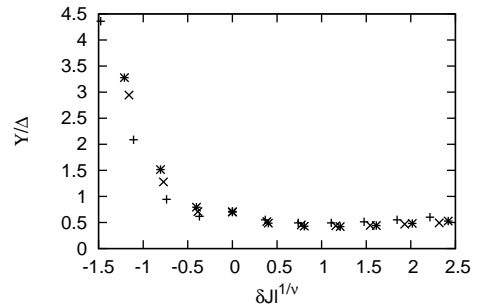}  }%
\caption{
\label{figure3}
The scaling plot of the amplitude ratio,
$\delta J \ell^{1/\nu}$-$\Upsilon (\delta J)/\Delta(-\delta J)$,
is presented
for the system sizes,
$N=30$ ($+$),
$32$ ($\times$), and
$34$ ($*$).
The scaling parameter $\nu$ is the same as that of Fig. \ref{figure2}.
In $\delta J \ell^{1/\nu} >0.5$, 
we observe a plateau with the height $\Upsilon/\Delta \sim 0.5$,
which indicates that the amplitude ratio takes a universal constant in proximity to the
critical point.
}
\end{figure}

To begin with,
we examine the criticality
of the paramagnetic gap $\Delta$,
which sets a fundamental energy scale
of this problem.
In Fig. \ref{figure2},
we present the 
scaling plot, 
$\delta J \ell ^{1/\nu}$-$\ell \Delta $,
for 
$N=30$ ($+$), 
$32$ ($\times$), and 
$34$ ($*$).
The paramagnetic gap 
$\Delta$ is calculated
by the formula
$\Delta=E_1-E_0$
with the ground-state energy $E_0$ ($E_1$)
within the total-magnetization
sector,
$S^z_{tot}=0$ ($1$).
It is an advantage of the exact diagonalization method
that such an excitation gap is calculated 
without resorting to the inverse Laplace transformation 
(see Appendix B of Ref. \cite{Gazit13}).

The scaling parameter (correlation-length critical exponent)
$\nu=0.7112$
is taken from the existing literatures 
\cite{Hasenbusch01,Campostrini02};
note that the criticality belongs \cite{Troyer98}
to the three-dimensional
Heisenberg universality class.
Hence, there is no adjustable fitting parameter
involved
in the scaling analysis.
Rather satisfactorily,
the scaled data obey the
finite-size scaling
for a considerably wide range of $\delta J$.
In Fig. \ref{figure2},
it is notable that
the paramagnetic gap closes (opens) in the (dis)ordered phase 
$\delta J>(<)0$.
In other words,
the critical point
(\ref{critical_point}) \cite{Troyer98}
as well as
the critical exponent
$\nu=0.7112$
\cite{Hasenbusch01,Campostrini02}
are supported by the present exact-diagonalization  analysis.
For such thermodynamic behavior, however,
the Monte Carlo method is more advantageous than the exact-diagonalization approach.
Hence, we do not pursue further details,
and turn our attention to the analysis of the transport properties.

We turn to the analysis of the amplitude ratio
$\Upsilon / \Delta$.
In Fig. \ref{figure3},
we present the scaling plot,
$\delta J \ell ^{1/\nu}$-$\Upsilon(\delta J)/\Delta (-\delta J)$,
for 
$N=30$ ($+$), 
$32$ ($\times$), and
$34$ ($*$).
Here, the scaling parameter $\nu$ is the same as that of Fig. \ref{figure2}.
The helicity modulus is calculated by the formula
\begin{equation}
\label{helicity_modulus}
\Upsilon = 
\frac{3}{2 \ell^2}
         \langle 0| K |0\rangle 
+\frac{3}{\ell^2}         
\left\langle 0
\left| \hat{J} 
\frac{{\cal P}}{ {\cal H}-E_0 } \hat{J}
\right|
0 \right\rangle
 .
\end{equation}
Here, the symbols
$E_0$ ($|0\rangle$), ${\cal P}$, $K$ and $\hat{J}$
denote
ground-state energy (eigenvector),
projection operator ${\cal P}=1-|0\rangle\langle0|$,
diamagnetic contribution,
and current operator, respectively;
in Appendix,
the
explicit formulas
for
$K$ and $\hat{J}$ are presented.
The overall prefactor $3/2$ is due to Ref. \cite{Sandvik97}.
The resolvent term 
(the second term of Eq. (\ref{helicity_modulus}))
was evaluated with the continued-fraction-expansion method \cite{Gagliano87}.
The continued-fraction-expansion method is essentially the same as
the Lanczos-tri-diagonalization sequence, and it is computationally less demanding.

In Fig. \ref{figure3},
we observe a plateau extending in
a considerably wide range of parameter 
$   \delta J \ell^{1/\nu} > 0.5$.
This plateau indicates that the amplitude ratio takes a constant value 
$\Upsilon / \Delta \sim 0.4$.
In a closer look,
we found
that the plateau takes an extremal point
$
\partial_{\delta J} ( \Upsilon (\delta J)/ \Delta (-\delta J)) |_{\delta J=\delta \bar{J}} = 0
$
at $\delta J= \delta \bar{J}$.
The plateau height at this point may serve a good indicator for $\Upsilon /\Delta$.

\begin{figure}
\resizebox{0.5\textwidth}{!}{%
\includegraphics{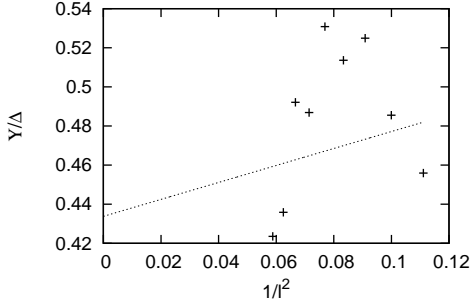}  }%
\caption{
\label{figure4}
The approximate amplitude ratio 
$\Upsilon / \Delta$ (\ref{approximate_amplitude_ratio1})
is plotted for $1/\ell^2$.
The least-squares fit yields an estimate 
$\Upsilon/\Delta=0.434(64)$
in the thermodynamic limit.
A bump around 
$1/\ell^2\approx0.82\dots( = 3.5^2)$ 
and a depression around $\approx 0.49 \dots (=1/4.5^2)$
are due to an artifact of the screw-boundary condition
\cite{Novotny90}.
A systematic error is considered in the text.
}
\end{figure}

In Fig. \ref{figure4},
we present the approximate amplitude ratio 
$\Upsilon / \Delta $
for $1/\ell^2$.
The approximate amplitude ratio denotes the plateau height
\begin{equation}
\label{approximate_amplitude_ratio1}
\Upsilon/\Delta=
\Upsilon (\delta J) / \Delta(-\delta J) |_{\delta J=\delta \bar{J}}
,
\end{equation}
for each system size.
The least-squares fit to these data yields an estimate 
$\Upsilon / \Delta =0.434(64)$
in the thermodynamic limit $\ell \to \infty$.
The data exhibit a wavy deviation;
the bump at 
$1/\ell^2 \approx 0.082 \dots (=1/3.5^2)$
and depression at
$\approx 0.049 \dots (=4.5^2)$ are due to an artifact of the screw-boundary condition
\cite{Novotny90}.
The wavy deviation amplitude 
appears to be $\sim 0.06$, which is bounded
by the above-mentioned 
least-squares-fit error $0.064$.
Accepting the uppermost value $0.07$
as an error margin,
we estimate the amplitude ratio as
\begin{equation}
\label{amplitude_ratio1}
\frac{\Upsilon}{\Delta}
\left(=\frac{\hbar}{2\pi \sigma_q L\Delta}
\right)=0.43(7)
.
\end{equation}
A comparison with 
the related studies is made in Sec. \ref{section2_3}.

Last, we address a remark as to the criticality of the bilayer
Heisenberg model (\ref{Hamiltonian})
as well as the scaling analyses of Figs. \ref{figure2} and \ref{figure3}.
The imaginary time and the spatial distance have the same scaling dimension.
Hence, the bilayer quantum model at
the ground state belongs to the three-dimensional universality class.
This mapping was confirmed by the analysis of Fig. \ref{figure2}.
The correlation-length critical exponent $\nu$ describes 
the singularity of the correlation length $\xi \sim \delta J^{-\nu}$.
Because the correlation length $\xi$ 
and the linear dimension of the cluster $\ell$
have the same scaling dimension,
the quantity $\delta J \ell^{1/\nu} $ should be scale-invariant.
This feature is the basis of the scaling analyses of Figs. \ref{figure2}
and \ref{figure3}, where the abscissa scale is set to
this scale-invariant parameter $\delta J \ell^{1/\nu} $.

\subsection{\label{section2_2}
Amplitude ratio $C/ L \sigma_q^2 (=\Upsilon \cdot 2\pi C/\hbar \sigma_q)$}

In this section,
we estimate the amplitude ratio 
$C/ L \sigma_q^2 (=\Upsilon \cdot 2\pi C / \hbar \sigma_q )$.
The capacitance $C$ is evaluated 
via the 
formula 
$C= \partial_k^2 \chi_\rho(k) / 2$ 
\cite{Gazit14,Nozieres58}
with the
charge-density-wave susceptibility $\chi_\rho(k)$ (see Appendix).
It is an advantage of the exact diagonalization method that the
capacitance is calculated directly at the ground state;
otherwise, the finite-temperature effect
has to be assessed carefully \cite{Gazit14s}.
In this section,
we survey the interaction subspace
$\delta J_2=0.15 \delta J$.

\begin{figure}
\resizebox{0.5\textwidth}{!}{%
\includegraphics{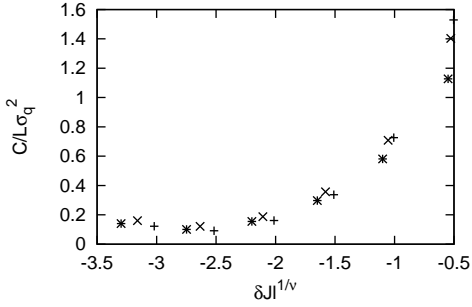}  }%
\caption{
\label{figure5}
The scaling plot of the amplitude ratio,
$\delta J \ell^{1/\nu}$-$C(\delta J)/L(-\delta J)\sigma_q^2$,
is presented
for the system sizes,
$N=30$ ($+$),
$32$ ($\times$), and
$34$ ($*$).
The scaling parameter $\nu$ is the same as that of Fig. 
\ref{figure2}.
In $\delta J \ell^{1/\nu}< -2$, we observe a plateau with the height 
$C/L\sigma_q^2\sim 0.1$.
}
\end{figure}

In Fig. \ref{figure5},
we present the scaling plot, 
$\delta J \ell^{1/\nu}$-$C(\delta J )/L( - \delta J) \sigma_q^2$,
for 
$N=30$ ($+$), 
$32$ ($\times$), and 
$34$ ($*$);
here, the scaling parameter $\nu$ is the same as that of Fig. \ref{figure2}.
We observe a plateau in the disordered phase $\delta J \ell ^{1/\nu}< -2$.
There appears an extremum point
$\partial_{\delta J} (C(\delta J)/L(-\delta J)\sigma_q^2)|_{\delta J=\delta \tilde{J}}=0$
at $\delta J=\delta \tilde{J}$.
The plateau height at this point
may provide a good indicator for $C/L\sigma_q^2$.

\begin{figure}
\resizebox{0.5\textwidth}{!}{%
\includegraphics{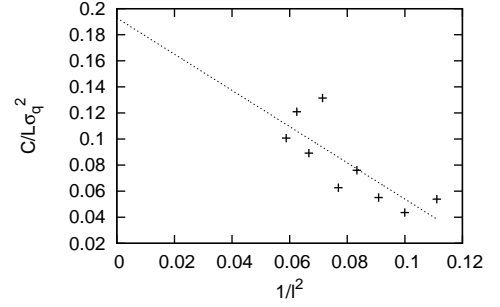}  }%
\caption{
\label{figure6}
The approximate amplitude ratio 
$C/L\sigma_q^2$ 
(\ref{approximate_amplitude_ratio2})
is plotted for $1/\ell^2$.
The least-squares fit yields an estimate 
$C/L\sigma_q^2=0.193(33)$
in the thermodynamic limit.
The intermittent bump
and shallow depression
around $1/\ell^2\approx 0.625(=1/4^2)$
and
$0.111\dots(=1/3^2)$, respectively,
are
due to an artifact of the screw-boundary condition \cite{Novotny90}.
A
possible systematic error is considered in the text.
}
\end{figure}

In Fig. \ref{figure6},
we present the approximate amplitude ratio
$C/L\sigma_q^2$
for $1/\ell^2$.
Here, the approximate amplitude ratio
denotes the plateau height
\begin{equation}
\label{approximate_amplitude_ratio2}
\frac{C}{L\sigma_q^2}=
\left.
\frac{C(\delta J)}{L(-\delta J)\sigma_q^2}
\right|_{\delta J=\delta \tilde{J} }
,
\end{equation}
for each system size.
The least-squares fit to these data yields
an estimate $C/L\sigma_q^2=0.193(33)$ in the thermodynamic limit.
The intermittent bump and shallow depression
around $1/\ell^2 \approx 0.625(=1/4^2)$
and 
$\approx 0.111\dots(=1/3^2)$, respectively,
 are due to an artifact of the
screw boundary condition \cite{Novotny90}.
Such wavy 
 deviation amplitude 
$\approx 0.03$ is bounded by
the above-mentioned least-squares-fit error
$0.033$.
Accepting the uppermost value $0.04$ as an error margin,
we estimate the amplitude ratio as
\begin{equation}
\label{amplitude_ratio2}
\frac{C}{L\sigma_q^2}
\left(
=\frac{\Upsilon\cdot 2\pi C}{\hbar\sigma_q}
\right)
=0.19(4)
.
\end{equation}

\subsection{\label{section2_3}
Set of
amplitude ratios 
$(\Upsilon/\Delta,\hbar \sigma_q/2\pi C\Delta,C/L\sigma_q^2)$: 
A brief overview}

The amplitude ratios, Eqs. (\ref{amplitude_ratio1})
and 
(\ref{amplitude_ratio2}),
immediately yield
yet another one
\begin{equation}
\label{amplitude_ratio3}
\hbar \sigma_q/(2\pi C \Delta)
=
2.3(6)
  .
\end{equation}
This amplitude ratio, in the O$(2)$ case, reduces to
$\rho_v/\Delta$,
which is dual to $\rho_s/\Delta$.
The above amplitude ratios,
Eqs. 
(\ref{amplitude_ratio1}),
(\ref{amplitude_ratio2}) and
(\ref{amplitude_ratio3}),
together with $m_H/\Delta$ 
\cite{Gazit13,Lohofer15,Rose15,Katan15,Nishiyama16}
almost fix the low-energy physics 
\cite{Chubukov94}
of the O$(3)$-symmetric
system
in proximity to the critical point.
The Higgs mode is hardly observable,
because it is smeared out by the Goldstone modes
\cite{Pekker15}.
Hence, it is significant to
fix the amplitude ratios
such as $m_H/ \rho_s$
quantitatively
in order to search for the 
(putative) Higgs branch 
hidden by
the Goldstone continuum.

This is a good position to address an overview 
on related studies; see Table \ref{table}.
First, the amplitude ratio 
$\Upsilon/\Delta$ was estimated with
the Blaizot-M\'endez-Galain-Wschebor  
(BMW) non-perturbative renormalization group (NPRG) method
as $\Upsilon/\Delta=0.401$ \cite{Rose15}.
Alternatively, with the derivative-expansion (DE)
NPRG method, the estimates,
$\Upsilon/\Delta=0.441$ \cite{Rose17}
and 
$0.3177$ \cite{Rancon13} were obtained.
According to
the Monte Carlo simulation \cite{Gazit13},
an estimate $\Upsilon/\Delta=0.34(1)$ was reported.
Our result (\ref{amplitude_ratio1})
supports recent NPRG studies,
$\Upsilon / \Delta=0.401$ \cite{Rose15} and 
$0.441$ \cite{Rose17};
as for the technical advantage of the former approach,
namely, the NPRG-BMW method,
we refer the reader to Ref. \cite{Blaizot06}.
Second,
 for $\hbar \sigma_q /2 \pi C \Delta$,
the NPRG-DE analysis \cite{Rose17} reported $1.98$.
Additionally, we draw reader's attention to its O$(2)$ counterpart
$1.98$ as well.
The results
for O$(2)$ and O$(3)$ seem to coincide with each other.
As a matter of fact,
according to the large-$N$ analysis \cite{Damle97},
this amplitude ratio converges to $12/2\pi=1.909\dots$ as $N\to \infty$.
Hence, it is suggested that 
the $N\to \infty$ consideration almost suffices
for the analysis of $\hbar \sigma_q/2\pi C \Delta$.
Last,
we turn to $C/L\sigma_q^2$.
Our result 
(\ref{amplitude_ratio2})
support
the NPRG-DE one $0.2226$ \cite{Rose17}.
These results indicates that 
a seemingly feasible relation $L/C\sigma_q^2\approx 1$
is not quite validated.
Hence,
so as to fix this amplitude ratio quantitatively,
 it is desirable to carry out
the non-perturbative analysis and
the brute-force calculation.

\section{\label{section3}
Summary and discussions}

The bilayer Heisenberg model (\ref{Hamiltonian})
was investigated
with
the exact diagonalization method,
which enables us to calculate the ground-state spectral 
and transport properties such as $\Delta$, $L$, and $C$ directly.
Thereby,
we shed light on its low-frequency conductivity
beside the critical point
(Fig. \ref{figure1}).
So far, the O$(2)$ case
has been investigated
with the aide of the
boson-vortex duality.
By means of the finite-size-scaling analysis for the
cluster with $N \le 34$ spins,
we obtained the amplitude ratios
$(\Upsilon/\Delta,\hbar\sigma_q/2\pi C \Delta,C/L\sigma_q^2)=
(0.43(7),2.3(6),0.19(4))$.
As for $\Upsilon/\Delta$,
with the NPRG and Monte Carlo methods,
there have been reported a number of estimates,
$0.441$ \cite{Rose17},
$0.401$ \cite{Rose15},
$0.3177$ \cite{Rancon13}, and
$0.34(1)$ \cite{Gazit13}.
Our result supports the
recent NPRG results $0.441$ \cite{Rose17}
and $0.401$ \cite{Rose15}.
Likewise, as for $\hbar\sigma_q/2\pi C \Delta$ and 
$C/L\sigma_q^2$,
our results agree with
those of
the recent NPRG study \cite{Rose17}, 
$1.98$ and $0.2226$, respectively.
The latter 
suggests that a seemingly feasible
relation
$C/L\sigma_q^2=1$ is not validated quantitatively.

As a matter of fact,
according to the preceeding computer-simulation analyses
\cite{Gazit13,Lohofer15,Nishiyama16}, there was reported an estimate
$m_H/\Delta \sim 2.2$-$2.7$,
which differs significantly from the mean-field value
$m_H/\Delta= \sqrt{2}$ \cite{Katan15}.
The spectral and transport properties 
seem to acquire
notable corrections
with respect to
those obtained through
the hand-waving arguments.
In this sense,
so as to fix the low-energy phenomenology 
of the O$(N)$-symmetric spectral and transport properties
\cite{Chubukov94},
the non-perturbative and brute-force approaches may be desirable.

\section*{Acknowledgment}
This work was supported by a Grant-in-Aid
for Scientific Research (C)
from Japan Society for the Promotion of Science
(Grant No. 25400402).

\begin{table}
\caption{
\label{table}
Preceeding results for O$(3)$ are summarized.
The non-perturbative renormalization group (NPRG) method
has a number of variants.
The abbreviations, DE and BMW,
denote derivative expansion and Blaizot M\'endez-Galain Wschebor,
respectively.
}
\label{table1}       
\begin{tabular}{llll}
\hline\noalign{\smallskip}
Amplitude ratios & $\frac{\Upsilon}{\Delta}(=\frac{\hbar}{2\pi \sigma_q L \Delta})$ & 
$ \frac{\hbar\sigma_q }{2\pi C\Delta}$ & 
    $\frac{ C }{L\sigma_q^2}$ \\
\noalign{\smallskip}\hline\noalign{\smallskip}
This work & 0.43(7) & 2.3(6) & 0.19(4) \\
NPRG-DE \cite{Rose17} & $0.441$ & $1.98$ & $0.2226$ \\
NPRG-BMW \cite{Rose15} & $0.401$ & & \\
NPRG-DE \cite{Rancon13} & $ 0.3177$ & & \\
Monte Carlo \cite{Gazit13} & $0.34(1)$ & & \\
\noalign{\smallskip}\hline
\end{tabular}
\end{table}

\appendix

\section*{Simulation algorithm: Screw-boundary condition}

In this paper, 
in order to implement the screw-boundary condition
\cite{Novotny90},
we adopted the simulation algorithm
as presented in
Eq. (A.1) of Ref. \cite{Nishiyama16}.
The screw-boundary condition
enables us
to
treat
a variety of system sizes
$N=18,20,\dots$ in a systematic manner.
The underlying idea behind this algorithm \cite{Novotny90}
is
that
an alignment of spins $S_i$ ($i=1,2,\dots$)
is wound up to form a toroidal coil,
which is equivalent to a rectangular cluster under the screw-boundary condition.
In the following, we present a number of extentions
in order to cope with the $J_2$ interaction and transport properties.
First, we need to incorporate the $J_2$ interaction.
For that purpose, we added
the term
$-J_2 \sum_{a=1}^2 \sum_{i=1}^{N/2} [
{\bf S}_{ai}(\ell+1)\cdot{\bf S}_{ai}
+{\bf S}_{ai}(\ell-1)\cdot{\bf S}_{ai}  ]$
with $\ell=\sqrt{N/2}$
to Eq. (A.1) of Ref. \cite{Nishiyama16}.   
Here, the symbol ${\bf S}_{ai}(\delta)$
denotes the $\delta$-shifted operator
${\bf S}_{ai}(\delta)= P^{-\delta}{\bf S}_{ai}P^\delta$
with the translation operator $P$ \cite{Novotny90}.
Second, the current operator $\hat{J}$ in Eq. 
(\ref{helicity_modulus})
is given by
\begin{eqnarray}
\hat{J} & =&
\frac{i J}{2} \sum_{a=1}^{2} \sum_{i=1}^{N/2}(S^+_{ai}(1)S^-_{ai}-
S^-_{ai}(1)S^+_{ai}) \nonumber \\
& & +
\frac{i J_2}{2} \sum_{a=1}^{2} \sum_{i=1}^{N/2}(S^+_{ai}(\ell+1)S^-_{ai}-
S^-_{ai}(\ell+1)S^+_{ai}) \nonumber \\
& & -
\frac{i J_2}{2} \sum_{a=1}^{2} \sum_{i=1}^{N/2}(S^+_{ai}(\ell-1)S^-_{ai}-
S^-_{ai}(\ell-1)S^+_{ai}) .
\end{eqnarray}
Likewise,
the diamagnetic part in Eq. (\ref{helicity_modulus}) is given by
\begin{eqnarray}
K & = &
\frac{ J}{2} \sum_{a=1}^{2} \sum_{i=1}^{N/2}(S^+_{ai}(1)S^-_{ai}+
S^-_{ai}(1)S^+_{ai}) \nonumber  \\
& &  +
\frac{ J_2}{2} \sum_{a=1}^{2} \sum_{i=1}^{N/2}(S^+_{ai}(\ell+1)S^-_{ai}+
S^-_{ai}(\ell+1)S^+_{ai}) \nonumber \\
& &  +
\frac{ J_2}{2} \sum_{a=1}^{2} \sum_{i=1}^{N/2}(S^+_{ai}(\ell-1)S^-_{ai}+
S^-_{ai}(\ell-1)S^+_{ai}) .
\end{eqnarray}
Last, we calculated the capacitance $C$ via the formula
$C=\frac{1}{\ell^2} \langle 0|N^\dagger_{k_1} ({\cal H}-E_0)^{-1}N_{k_1}|0\rangle / k_1^2$
with $N_k=\sum_{a=1}^2 \sum_{j=1}^{N/2} e^{ikj} S^z_{aj}$ and $k_1=2\pi \div \frac{N}{2}$
\cite{Gazit14,Nozieres58,Gazit14s}.

%

%
%
%

\begin{thebibliography}{}
%
%


\bibitem{Gazit13}
S. Gazit, D. Podolsky, A. Auerbach, and
 D. P. Arovas,
Phys. Rev. B {\bf 88}, 235108 (2013).


\bibitem{Chubukov94}
A. V. Chubukov, S. Sachdev, and J. Ye,
Phys. Rev. B {\bf 49}, 11919 (1994).



\bibitem{Rose17}
F. Rose and N. Dupuis, Phys. Rev. B {\bf 95}, 014513 (2017). 

\bibitem{Lohofer15}
M. Loh\"ofer, T. Coletta, D. G. Joshi, F. F. Assaad, M. Vojta,
S. Wessel, and F. Mila, Phys. Rev. B {\bf 92}, 245137 (2015).
\bibitem{Rose15}
F. Rose, F. L\'eonard, and N. Dupuis, Phys. Rev. B {\bf 91}, 224501 (2015).
\bibitem{Katan15}
Y. T. Katan and D. Podolsky, Phys. Rev. B {\bf 91}, 075132 (2015).
\bibitem{Nishiyama16}Y. Nishiyama, Eur. Phys. J. B {\bf 89}, 31 (2016).



\bibitem{Rancon14}A. Ran\c{c}on and N. Dupuis, Phys. Rev. B {\bf 89},
180501 (2014).






\bibitem{Stone78}M. Stone and P. R. Thomas, 
Phys. Rev. Lett. {\bf 41}, 351 (1978). 

\bibitem{Fisher89}M. P. A. Fisher and D. H. Lee,  
Phys. Rev. B {\bf 39}, 2756 (1989).

\bibitem{Wen90}X. G. Wen and A. Zee, 
Int. J. Mod. Phys. B {\bf 04}, 437 (1990).





\bibitem{Gazit14}
S. Gazit, D. Podolsky and A. Auerbach,
Phys. Rev. Lett. {\bf 113}, 240601 (2014). 






\bibitem{Corson99}
J. Corson, R. Mallozz, J. Orenstein, J. N. Eckstein,
and  I. Bozovic,
Nature {\bf 398}, 221 (1999).

\bibitem{Crane07}
R. W. Crane, N. P. Armitage, A. Johansson, G. Sambandamurthy,
D. Shahar, and G. Gr\"uner,
Phys. Rev. B {\bf 75}, 094506 (2007).

\bibitem{Sherson10}
J. F. Sherson, C. Eeitenberg, M. Endres,
 M. Cheneau, I. Bloch, and S. Kuhr,
Nature {\bf 467}, 68 (2010).






%
\bibitem{Troyer98}
M. Troyer and S. Sachdev, Phys. Rev. Lett. {\bf 81}, 5418 (1998).
%
\bibitem{Hasenbusch01}
M. Hasenbusch, J. Phys. A {\bf 34}, 8221 (2001).
\bibitem{Campostrini02}
M. Campostrini, M. Hasenbusch, A. Pelissetto,
P. Rossi, and E. Vicari, Phys. Rev. E {\bf 65}, 144520 (2002).

\bibitem{Rancon13}
A. Ran\c{c}son, O. Kdio, N. Dupuis, and P. Lecheminant, 
Phys. Rev. E {\bf 88}, 012113 (2013). 


\bibitem{Sentef07}
M. Sentef, M. Kollar, and A. P. Kampf, Phys. Rev. B {\bf 75}, 214403 (2007).
\bibitem{Pardini08}
T. Pardini, R. R. P. Singh, A. Katanin, and O. P. Sushkov,
Phys. Rev. B {\bf 78}, 024439 (2008).
\bibitem{Kubo13}
Y. Kubo and S. Kurihara, J. Phys. Soc. Japan {\bf 82}, 113601 (2013).
\bibitem{Chen13}
Z. Chen, T. Datta, and D.-X. Yao, Eur. Phys. J. B {\bf 86}, 63 (2013).



%
\bibitem{Novotny90}M. A. Novotny, J. Appl. Phys. {\bf 67}, 5448 (1990).





\bibitem{Sandvik97}
A.W. Sandvik, Phys. Rev. B {\bf 56}, 11678 (1997).


\bibitem{Gagliano87}
E. R. Gagliano and C. A. Balseiro,
Phys. Rev. Lett. {\bf 59}, 2999 (1987).




\bibitem{Nozieres58}
P. Nozi\`eres and D. Pines,
Nuovo Cim. {\bf 9} (1958) 470.




\bibitem{Gazit14s}
See Supplemental Material of Ref. \cite{Gazit14}.


\bibitem{Pekker15}D. Pekker and C.M. Varma, Annual Rev. Condens. Matter
Phys. {\bf 6}, 269 (2015).






\bibitem{Blaizot06}
J.-P. Blaizot, R. M\'endez-Galain, and
N. Wschebor, Phys. Lett. B {\bf 632}, 571 (2006).

\bibitem{Damle97}
K. Damle and S. Sachdev,
Phys. Rev. B {\bf 56}, 8714 (1997).




\end{thebibliography}
%

\end{document}